# Atomic-scale fragmentation and collapse of antiferromagnetic order in a doped Mott insulator

He Zhao[1], Sujit Manna[2,3,4], Zach Porter[5], Xiang Chen[5], Andrew Uzdejczyk[1], Jagadeesh Moodera[2,3,4], Ziqiang Wang[1], Stephen D. Wilson[5] and Ilija Zeljkovic[1,*]

**Affiliations:**

[1] Department of Physics, Boston College, 140 Commonwealth Ave, Chestnut Hill, MA 02467

[2] Francis Bitter Magnet Laboratory, Massachusetts Institute of Technology, Cambridge, MA 02139, USA

[3] Department of Physics, Massachusetts Institute of Technology, Cambridge, MA 02139, USA.

[4] Plasma Science and Fusion Center, Massachusetts Institute of Technology, Cambridge, MA 02139, USA

[5] Materials Department, University of California, Santa Barbara, CA 93106, USA.

[*]Correspondence to: ilija.zeljkovic@bc.edu

**Disentangling the relationship between the insulating state with a charge gap and the magnetic order in an antiferromagnetic (AF) Mott insulator remains difficult due to inherent phase separation as the Mott state is perturbed. Measuring magnetic and electronic properties at the atomic length scales would provide crucial insight, but this is yet to be experimentally achieved. Here we use spectroscopic-imaging spin-polarized scanning tunneling microscopy (SP-STM) to visualize periodic spin-resolved modulations originating from the AF order in a relativistic Mott insulator $Sr_2IrO_4$, and study these as a function of doping. We find that near insulator-to-metal transition (IMT), the long-range AF order melts into a fragmented state with short-range AF correlations. Crucially, we discover that the short-range AF order is locally uncorrelated with the observed spectral gap magnitude. This strongly suggests that short range AF correlations are unlikely to be the culprit behind inhomogeneous gap closing and the emergence of pseudogap regions near IMT. Our work establishes SP-STM as a powerful tool for revealing atomic-scale magnetic information in complex oxides.**

A Mott insulator, characterized by localization of electrons due to strong electron-electron interactions (*1*), is typically accompanied by magnetic ordering (*2–4*). The antiferromagnetic (AF)

Mott insulator has been of particular interest as the low-temperature ground state of numerous transition metal oxides (*3–6*), most notably cuprate high-temperature superconductors. Charge carrier doping can gradually suppress this insulating state, a process theoretically expected to lead to inherent phase separation (*7–9*). Consistent with this picture of doped Mott insulators, spatially inhomogeneous electronic states, varying on nanometer length scales, have been observed in several systems (*10–15*). However, a major experimental failing over the past few decades has been the inability to measure how the AF order evolves over the same length scales, and what is its relationship to the single-particle gap (charge gap) closing as the Mott state is tuned away from half filling. Spin-polarized scanning tunneling microscopy (SP-STM) is one of the few available techniques that can in principle be used for atomic-scale imaging of underlying magnetic orders (*16*). However, it has so far mostly been applied to magnetic nanostructures (*16–21*) and Fe-based systems (*22–24*).

$Sr_2IrO_4$ (Sr-214) is a quasi-2D layered oxide, in which strong spin-orbit coupling and on-site Coulomb repulsion $U$ lead to the formation of an effective $J_{eff}=1/2$ Mott state (*4*, *25*). The Mott state is accompanied by an in-plane AF ordering of Ir spins below ~ 240 K (Fig. 1(a,b)) and a cooperative in-plane rotation of the $IrO_6$ octahedra (*4*), both of which lead to doubling of the primitive unit cell of Ir cations (Fig. 1c). Interestingly, Sr-214 doped by electrons is thought to be a nearly perfect analogue to hole-doped cuprates (*4*, *15*). It exhibits many of the same electronic phenomena emerging with doping, such as a pseudogap (*12*, *15*, *26–28*), Fermi arcs (*29*) and an incommensurate spin density wave (*30*). Although superconductivity is yet to be achieved, it has been theoretically predicted to occur at higher electron densities (*31*), if those were to be experimentally achieved. In contrast to cuprates where the AF Mott state is suppressed by ~0.05 holes per Cu site, the AF Mott state in Sr-214 is only partially quenched even at the highest achievable doping of ~ 0.12 electrons per Ir site (*14*, *32*). Moreover, compared to a relatively large charge gap in insulating cuprates (*11*, *33*), a smaller gap in lightly-doped, insulating Sr-214 allows tunneling experiments to be more easily performed (*4*, *34*). As such, this system presents an ideal platform to investigate the interplay of magnetism and electronic correlations within the lightly-doped Mott state. Local spectroscopy measurements of electron-doped Sr-214 have provided a glimpse into a spatially inhomogeneous electronic state near $x_c$, with metallic regions dispersed within an insulating background at the highest available doping (*12*, *14*). However, a crucial question remains – how does the underlying magnetic order evolve towards IMT and what is its relationship to the inhomogeneous electronic state?

To investigate this, we synthesize and study single crystals of $(Sr_{1-x}La_x)_2IrO_4$ (La-doped Sr-214), in which each La substitution contributes ~1 electron to the system. We explore samples in two distinct doping regimes – one deep in the Mott phase with La concentration $x \sim 0.02$ and another near the IMT with nominal $x \sim 0.05$. Consistent with previous experiments (*12*, *14*), a typical STM topograph of our La-doped Sr-214 (which inevitably contains both electronic and structural information) acquired with a spin-averaged W tip reveals a square lattice of Sr atoms in the topmost SrO plane with the lattice constant $a_0 \sim 3.9$Å (Fig. 1d). In addition to the atomic Bragg peaks $\mathbf{Q_x}=(1,0)$ and $\mathbf{Q_y}=(0,1)$ (we hereafter define reciprocal lattice vector $2\pi/a_0 = 1$), Fourier transform (FT) of the STM topograph also shows peaks at $\mathbf{Q_a}=(-\frac{1}{2},\frac{1}{2})$ and $\mathbf{Q_b}=(\frac{1}{2},\frac{1}{2})$ (Fig. 1e), which are hypothesized to be a consequence of the staggered rotation of $IrO_6$ octahedra (*15*). The AF ordering in Sr-214, which has been well-established by neutron scattering experiments (*14*, *35*), is also spatially modulated with $\mathbf{Q_a}$ and $\mathbf{Q_b}$ wave vectors (Fig. 1(b,c)), but its signal would be difficult to detect by conventional spin-averaged tunneling using a W tip. In this work, to achieve spin-resolved imaging, we use spin-polarized bulk Chromium (Cr) tips characterized on the surface of antiferromagnet $Fe_{1+y}Te$ (Methods, Supplementary Information 1).

We start with spectroscopic characterization of the $x\sim0.02$ sample. Consistent with previous work (*12*), our STM $dI/dV$ spectra reveal a large insulating U-shaped gap across the entire field of view, which locally decreases in proximity to La dopants (Fig. 1f). Next, we focus on measuring the magnetic signal at $\mathbf{Q_a}=(-\frac{1}{2},\frac{1}{2})$ and $\mathbf{Q_b}=(\frac{1}{2},\frac{1}{2})$. To achieve this, we use a Cr tip to acquire STM topographs $T(\mathbf{r}, \mathbf{B_i})$ of the same region of the sample under varying magnetic field $\mathbf{B_i} \parallel$ c-axis, while keeping all other experimental conditions the same. Typically, a Cr tip has a canted magnetization direction $\mathbf{M}_{tip}$, with both in-plane and out-of-plane components even at zero magnetic field (*16*). By applying an external field $\mathbf{B_i}$, $\mathbf{M}_{tip}$ tilts towards the direction of $\mathbf{B_i}$. As the direction of $\mathbf{M}_{tip}$ changes, its overlap with the AF-ordered spins in the sample also changes, which in turn modulates the measured amplitude of the electronic signal at $\mathbf{Q_a}$ and $\mathbf{Q_b}$. The FTs of STM topographs obtained using a spin-resolved STM tip show a systematic evolution of $\mathbf{Q_a}$ and $\mathbf{Q_b}$ amplitudes (Fig. 2g). Since nearest neighbor exchange interactions (*36*) are much stronger than the Zeeman energy at the relatively modest range of fields used in our experiments, we expect no significant change of the AF order of the sample. Thus, the change in $Q_a$ and $Q_b$ occurs due to the evolution of the direction of $M_{tip}$. We note that the FTs of STM topographs of the same sample acquired using a spin-averaged W tip are nearly identical to one another, and that all FT peaks remain approximately constant upon changing **B** (Fig. 2h, Supplementary Information 2). This demonstrates that the change in $Q_a$ and $Q_b$ with magnetic field observed with a Cr tip (Fig. 2g) is primarily due to spatially-modulated spin-polarized electron tunneling.

To visualize the spatial distribution of the measured spin-polarized signal, we first align topographic images acquired at two different fields T(**r**, $B_1$) and T(**r**, $B_2$) by applying the Lawler-Fujita drift-correction algorithm (*37*) (Fig. 2a,b) and by using La-dopants as position markers. As discussed previously, changing the magnetic field from $B_1$ to $B_2$ is only expected to affect the direction of $M_{tip}$ (typically both $B_1$ and $B_2$ have the same magnitude but are applied along opposite directions). Then, similarly to the procedure in Ref. (*22*), the difference M(**r**)=T(**r**,$B_1$)-T(**r**,$B_2$) yields the spin-resolved magnetic contrast (Fig. 2d), while the average of the two images N(**r**) = (T(**r**,$B_1$) + T(**r**,$B_2$))/2 primarily contains non-magnetic electronic information (Fig. 2c). Remarkably, the spin-resolved magnetic contrast image M(**r**) of the *x*~0.02 sample displays a prominent bi-directional modulation pattern, which is notably absent in the equivalent image obtained on the same sample using a spin-averaged W tip (Fig. S4). The local amplitude of the modulations appears to be homogeneous across the image, unrelated to the local density of La dopants (Fig. 2d). The FT of M(**r**) indicates that this is a commensurate order, oriented at 45 degrees with respect to the Sr lattice, with wave vectors narrowly centered at a single pixel at exactly $Q_a=(-\frac{1}{2},\frac{1}{2})$ and $Q_b=(\frac{1}{2},\frac{1}{2})$ (Fig. 2f). This is consistent with the well-established long-range AF order detected by neutron diffraction at this La concentration (*14*, *30*). We note that an STM topograph contains local density-of-states information from the Fermi level to $eV_{sample}$ ($V_{sample}$ ~ 500-700 meV used in this work). Importantly, $J_{eff}$= 1/2 band is expected to be the only band present in this energy range, as $J_{eff}$= 3/2 is pushed below the Fermi level and other bands are expected at much higher energies (*38*, *39*). This strongly suggests that spin-resolved modulations in M(**r**) are predominantly associated with $J_{eff}$= 1/2 band itself.

Next, we repeat the measurement on the *x* ~ 0.05 sample (Fig. 3), which is expected to be close to the IMT (*14*). We find that M(**r**) image also exhibits two-dimensional modulations (Fig. 3d). However, in stark contrast to lower doping, we can clearly observe domain walls (solid line in Fig.

3d, Fig. S7), as well as more diffused FT peaks centered at $\mathbf{Q_a} = (-\frac{1}{2}, \frac{1}{2})$ and $\mathbf{Q_b} = (\frac{1}{2}, \frac{1}{2})$, which are no longer contained to a single pixel (Fig. 3f). Furthermore, no new peaks are observable. Interestingly, in some regions of the sample, the modulations appear weaker and even disappear (dashed regions in Fig. 3d), which suggests a wide distribution of local AF strengths. The observed fragmented state is consistent with a short-range AF order inferred from magnetization measurements at this approximate doping (*14*). Given that Sr-214 at even higher electron doping should be paramagnetic (*14*), it is conceivable that the non-ordered regions in M(**r**) in Fig. 3 could also be paramagnetic. The auto-correlation of M(**r**) acquired over a larger region of the $x \sim 0.05$ sample (Fig. 4c) gives the correlation length-scale of $\xi_{AF} \sim 5$ nm. We can visualize the local strength of magnetic correlations by constructing an amplitude map (Fig. 4d), which highlights the fragmented nature of the state, with ordered puddles varying in size from ~10 to >100 IrO$_2$ plaquettes. We note that intra-unit-cell positions of the modulation peaks in M(**r**) are centered somewhat away from the Ir site (Fig. 4c), which we attribute to the electron tunneling path from the IrO$_2$ plane through the surface SrO layer involving a complex overlap of orbitals and/or possibly O bands mixing into the $J_{eff} = 1/2$ band. To exclude the effects of a particular STM tip and experimental parameters on the observation of the short-range order, we show M(**r**) images of multiple different samples, acquired using different bulk Cr tips and magnitudes of magnetic fields that exhibit qualitatively the same behavior (Supplementary Information 2).

Finally, we proceed to investigate the relationship between the spatial distribution of the spin-resolved magnetic contrast M(**r**) and the underlying electronic structure. In accordance with previous spectroscopic measurements of the $x \sim 0.05$ sample (*12*, *14*), we observe two main types of *dI/dV* spectra: U-shaped Mott like spectra similar to those at low doping (blue curve in Fig. 1g) and V-shaped Mott/pseudogap like spectra with a smaller overall gap (gray curve in Fig. 1g). To quantify local electronic structure, we acquire *dI/dV* spectra on a densely spaced pixel grid and create a map of the approximate gap magnitude Δ(**r**) over the same region of the sample as M(**r**) (Fig. 4b, Supplementary Information 3). Theoretical calculations indicate that opening of the gap in the insulating state should scale with the strength of the AF ordering (*38*). In the scenario where the antiferromagnetic state is the only ordered state present, the AF-ordered regions would be expected to exhibit larger Δ(**r**) and Mott like spectra, but as the AF order locally weakens and collapses, the gap should also shrink. Therefore, the magnitude of Δ(**r**) should strongly correlate with the amplitude of M(**r**). Surprisingly however, we find that the two observables exhibit no cross-correlation (Fig. 4f). This conclusion can also be confirmed by a visual inspection of Figs. 4(a-d), where we can clearly distinguish patches hosting short-range spin-polarized modulations with either a small V-shaped gap (solid blue square) or a large Mott like gap (solid black square). Similarly, we can also see regions where spin-polarized modulations are absent, but the sample shows either a V-shaped (dashed blue square) or a Mott like spectrum (dashed black square).

Our experiments suggest that short-range AF correlations by themselves are unlikely to be the culprit behind the inhomogeneous gap closing near IMT. The residual gaps near IMT could in principle be explained by a pure Mott charge gap in the weak Mott state, but we deem this possibility unlikely given the high level of electron doping and local proximity to dopants. This brings a surprising possibility suggestive of another, "hidden" ordered state that may be contributing to the gap modulations, such as the magneto-electric loop-current order reported in the parent and hole-doped Sr-214 (*40*), or the spin-orbit density wave (SODW) consisting of counter-propagating spin currents theoretically proposed to occur in electron-doped Sr-214 (*38*). The latter is an intriguing possibility as it is found to strongly modulate the gap magnitude at the X point by $\sim \pm 100$ meV, which may explain the surprising relationship between the electronic structure and the AF order we found here (Fig. 4). It remains to be seen what parameter locally controls the patchy distribution of the spin-resolved magnetic contrast M(**r**), as the observed domains do not seem to exhibit any obvious correlation with the spatial distribution of La dopants in the top SrO plane (Fig. 3d). It may be possible that La dopants in the SrO plane below the $IrO_2$ layer have some effect on the local amplitude of M(**r**), but these could not be detected in our experiments. To elucidate if the domains are pinned by the presence (or absence) of disorder, we propose acquiring M(**r**) images of the same region of the sample, before and after warming up through the paramagnetic transition. If it is found that these domains shift upon thermal cycling, this may present an exciting opportunity to investigate the effect of moving an AF domain on the local electronic structure. Our work provides a new insight into how an AF Mott insulator melts with charge carrier doping, and establishes SP-STM as a powerful tool to probe magnetism in complex oxides, where spin, charge, and orbital degrees of freedom intertwine at the nanoscale.

## Methods

**Single crystal growth.** La-doped Sr-214 single crystals were grown using flux growth method (*14*). Starting powders used were $SrCO_3$ (99.99%, Alfa Aesar), $La_2O_3$ (99.99%, Alfa Aesar), $IrO_2$ (99.99%, Alfa Aesar), and anhydrous $SrCl_2$ (99.5%, Alfa Aesar) mixed in a $2(1-x)$: $x$: 1: 6 molar ratio, where $x$ is the nominal La concentration. They were fully ground, mixed and placed inside a Platinum (Pt) crucible, capped by a Pt lid, and further protected by an outer alumina crucible. Mixtures were heated slowly to 1380 °C, soaked for ~5-10 hours, slowly cooled to 850 °C over the course of 120 hours and then furnace cooled to room temperature over ~5 hours. Single crystals used in the work were then obtained after removing excess flux with deionized water.

**SP-STM measurements.** La-doped Sr-214 samples were cleaved in ultra-high vacuum at ~80 K and immediately inserted into the STM head. All STM data was acquired at the base temperature of ~4.5 K using a commercial Unisoku USM1300 system. Spectroscopic measurements were taken using a standard lock-in technique at 915 Hz frequency and varying bias excitation as detailed in the figure captions. The STM tips used were home-made, either chemically etched bulk W tips or bulk Cr tips, as indicated in the figure captions. To obtain an atomically sharp Cr tip, Cr wire (obtained by cutting a ~1 mm thick Cr sheet into strips, each one with ~1 x ~1 mm cross-section) was etched in 2 mol/L NaOH solution over the course of 30 minutes, similar to the procedure described in Ref. (*41*). Tip sharpness was first evaluated under optical microscope, before the tip was annealed in UHV and inserted into the STM, where its sharpness and spin-polarization was assessed on the surface of UHV-cleaved antiferromagnet $Fe_{1+y}Te$ (Supplementary Information 1).

## References

1. N. F. Mott, The Basis of the Electron Theory of Metals, with Special Reference to the Transition Metals. *Proc. Phys. Soc. A.* **62**, 416–422 (1949).
2. M. Imada, A. Fujimori, Y. Tokura, Metal-insulator transitions. *Rev. Mod. Phys.* **70**, 1039–1263 (1998).
3. P. A. Lee, N. Nagaosa, X. G. Wen, Doping a Mott insulator: Physics of high-temperature superconductivity. *Rev. Mod. Phys.* **78**, 17–85 (2006).
4. G. Cao, P. Schlottmann, The challenge of spin-orbit-tuned ground states in iridates: A key issues review. *Reports Prog. Phys.* **81**, 1–58 (2018).

5.	M. Imada, A. Fujimori, Y. Tokura, Metal-insulator transitions. *Rev. Mod. Phys.* **70**, 1039–1263 (1998).

6.	J. G. Rau, E. K.-H. Lee, H.-Y. Kee, Spin-Orbit Physics Giving Rise to Novel Phases in Correlated Systems: Iridates and Related Materials. *Annu. Rev. Condens. Matter Phys.* **7**, 195–221 (2016).

7.	P. B. Visscher, Phase separation instability in the Hubbard model. *Phys. Rev. B.* **10**, 943–945 (1974).

8.	V. Emery, S. Kivelson, H. Lin, Phase separation in the *t-J* model. *Phys. Rev. Lett.* **64**, 475–478 (1990).

9.	C. H. Yee, L. Balents, Phase separation in doped Mott insulators. *Phys. Rev. X.* **5**, 1–5 (2015).

10.	M. M. Qazilbash *et al*. Mott Transition in $VO_2$ Revealed by Infrared Spectroscopy and Nano-Imaging. *Science* **318**, 1750–1753 (2007).

11.	P. Cai *et al*. Visualizing the evolution from the Mott insulator to a charge-ordered insulator in lightly doped cuprates. *Nat. Phys.* **12**, 1047–1051 (2016).

12.	I. Battisti *et al*. Universality of pseudogap and emergent order in lightly doped Mott insulators. *Nat. Phys.* **13**, 21–25 (2016).

13.	C. Dhital *et al*. Carrier localization and electronic phase separation in a doped spin-orbit-driven Mott phase in $Sr_3(Ir_{1-x}Ru_x)_2O_7$. *Nat. Commun.* **5**, 3377 (2014).

14.	X. Chen *et al*. Influence of electron doping on the ground state of $(Sr_{1-x}La_x)_2IrO_4$. *Phys. Rev. B*. **92**, 075125 (2015).

15.	Y. J. Yan *et al*. Electron-doped $Sr_2IrO_4$: An analogue of hole-doped cuprate superconductors demonstrated by scanning tunneling microscopy. *Phys. Rev. X.* **5**, 1–7 (2015).

16.	R. Wiesendanger, Spin mapping at the nanoscale and atomic scale. *Rev. Mod. Phys.* **81**, 1495–1550 (2009).

17.	I. V. Shvets *et al*. Progress towards spin-polarized scanning tunneling microscopy. *J. Appl. Phys.* **71**, 5489 (1992).

18.	S. Jeon *et al*. Distinguishing a Majorana zero mode using spin-resolved measurements. *Science* **358**, 772–776 (2017).

19.	F. D. D. Natterer *et al*. Reading and writing single-atom magnets. *Nature* **543**, 226–228 (2017).

20. C. F. F. Hirjibehedin, C. P. P. Lutz, A. J. J. Heinrich, Spin Coupling in Engineered Atomic Structures. *Science* **312**, 1021–1024 (2006).

21. S. Loth, S. Baumann, C. P. P. Lutz, D. M. M. Eigler, A. J. J. Heinrich, Bistability in Atomic-Scale Antiferromagnets. *Science* **335**, 196–199 (2012).

22. M. Enayat *et al*. Real-space imaging of the atomic-scale magnetic structure of $Fe_{1+y}Te$. *Science* **345**, 653–656 (2014).

23. S. Manna *et al*. Interfacial superconductivity in a bi-collinear antiferromagnetically ordered FeTe monolayer on a topological insulator. *Nat. Commun.* **8**, 14074 (2017).

24. S. Choi *et al*. Switching Magnetism and Superconductivity with Spin-Polarized Current in Iron-Based Superconductor. *Phys. Rev. Lett.* **119**, 227001 (2017).

25. B. Kim *et al*. Novel $J_{eff}$=1/2 Mott State Induced by Relativistic Spin-Orbit Coupling in $Sr_2IrO_4$. *Phys. Rev. Lett*. **101**, 076402 (2008).

26. A. de la Torre *et al*. Collapse of the Mott Gap and Emergence of a Nodal Liquid in Lightly Doped $Sr_2IrO_4$. *Phys. Rev. Lett.* **115**, 176402 (2015).

27. Y. Cao *et al*. Hallmarks of the Mott-metal crossover in the hole-doped pseudospin-1/2 Mott insulator $Sr_2IrO_4$. *Nat. Commun.* **7**, 11367 (2016).

28. Y. K. Kim, N. H. Sung, J. D. Denlinger, B. J. Kim, Observation of a d-wave gap in electron-doped $Sr_2IrO_4$. *Nat. Phys.* **12**, 37–41 (2016).

29. Y. K. Kim *et al*. Fermi arcs in a doped pseudospin-1/2 Heisenberg antiferromagnet. *Science* **345**, 187–190 (2014).

30. X. Chen *et al*. Unidirectional spin density wave state in metallic $(Sr_{1-x}La_x)_2IrO_4$. *Nat. Commun.* **9**, 103 (2018).

31. F. Wang, T. Senthil, Twisted hubbard model for Sr2IrO4: Magnetism and possible high temperature superconductivity. *Phys. Rev. Lett*. **106**, 136402 (2011).

32. M. Ge *et al*. Lattice-driven magnetoresistivity and metal-insulator transition in single-layered iridates. *Phys. Rev. B*. **84**, 100402 (2011).

33. H. Zhao *et al*. Charge-stripe crystal phase in an insulating cuprate. *Nat. Mater.* **18**, 103–107 (2019).

34. J. M. Guevara *et al*. Spin-polaron ladder spectrum of the spin-orbit-induced Mott insulator $Sr_2IrO_4$ probed by scanning tunneling spectroscopy. arxiv:1802.10028.

35. F. Ye *et al*. Magnetic and crystal structures of $Sr_2IrO_4$: A neutron diffraction study. *Phys. Rev. B.* **87**, 140406 (2013).

36. J. Kim *et al*. Magnetic Excitation Spectra of $Sr_2IrO_4$ Probed by Resonant Inelastic X-Ray Scattering: Establishing Links to Cuprate Superconductors. *Phys. Rev. Lett.* **108**, 177003 (2012).

37. M. J. Lawler *et al*. Intra-unit-cell electronic nematicity of the high-$T_c$ copper-oxide pseudogap states. *Nature* **466**, 347–351 (2010).

38. S. Zhou, K. Jiang, H. Chen, Z. Wang, Correlation effects and hidden spin-orbit entangled electronic order in parent and electron-doped iridates $Sr_2IrO_4$. *Phys. Rev. X.* **7**, 1–9 (2017).

39. I. V. Solovyev, V. V. Mazurenko, A. A. Katanin, Validity and limitations of the superexchange model for the magnetic properties of $Sr_2IrO_4$ and $Ba_2IrO_4$ mediated by the strong spin-orbit coupling. *Phys. Rev. B.* **92**, 1–19 (2015).

40. L. Zhao *et al*. Evidence of an odd-parity hidden order in a spin–orbit coupled correlated iridate. *Nat. Phys*. **12**, 32–36 (2016).

41. D. Huang, S. Liu, I. Zeljkovic, J. F. Mitchell, J. E. Hoffman, Etching of Cr tips for scanning tunneling microscopy of cleavable oxides. *Rev. Sci. Instrum*. **88**, 023705 (2017).

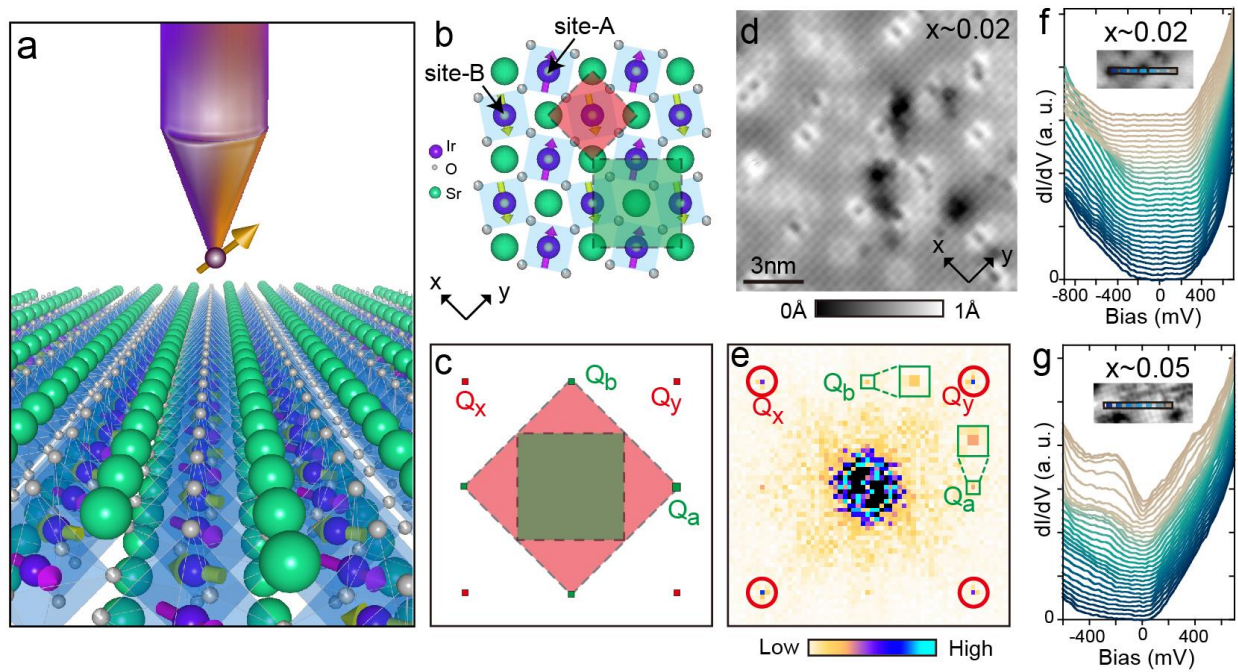

**Figure 1. Measurement schematic, crystal structure and basic electronic characterization.**
(a) Schematic of the spin-polarized scanning tunneling microscopy (SP-STM) measurement of $(Sr_{1-x}La_x)_2IrO_4$ (La-doped Sr-214). For visualization purposes, the arrow at the apex denotes the spin-polarization of the tip. (b) Crystal structure of Sr-214 in the *xy* plane. Both Sr lattice unit cell (pink square) and the superstructure unit cell due to $IrO_6$ octahedral rotation (green square, also magnetic unit cell) are shown. The arrows schematically depict spin momenta of Ir atoms. (c) The expected Fourier transform (FT) image of the Sr-214 STM topograph. Pink and green squares correspond to the 1$^{st}$ Brillouin zone of the Sr lattice and the superstructure, respectively. $\mathbf{Q_x}$ and $\mathbf{Q_y}$ denote the atomic Bragg peaks of the Sr lattice, while $\mathbf{Q_a}$ and $\mathbf{Q_b}$ in principle contain contributions from the octahedral rotation and AF ordering. (d) Representative STM topograph of nominal *x*~0.02 La-doped Sr-214 sample and (e) associated FT obtained using a spin-averaged W tip. La substitutions are visible as bright squares in the STM topograph in (d). A series of *dI/dV* spectra acquired on the surface of (f) *x*~0.02 and (g) *x*~0.05 La-doped Sr-214, along the line shown in each inset. STM setup conditions: (d) $V_{sample}$= 700 mV, $I_{set}$ = 100 pA, B = 0 T; (f) $V_{sample}$= 700 mV, $I_{set}$ = 100 pA, $V_{exc}$= 5 mV (zero-to-peak), B = 0 T; (g) $V_{sample}$= 700 mV, $I_{set}$ = 250 pA, $V_{exc}$= 8 mV (zero-to-peak), B = 0 T.

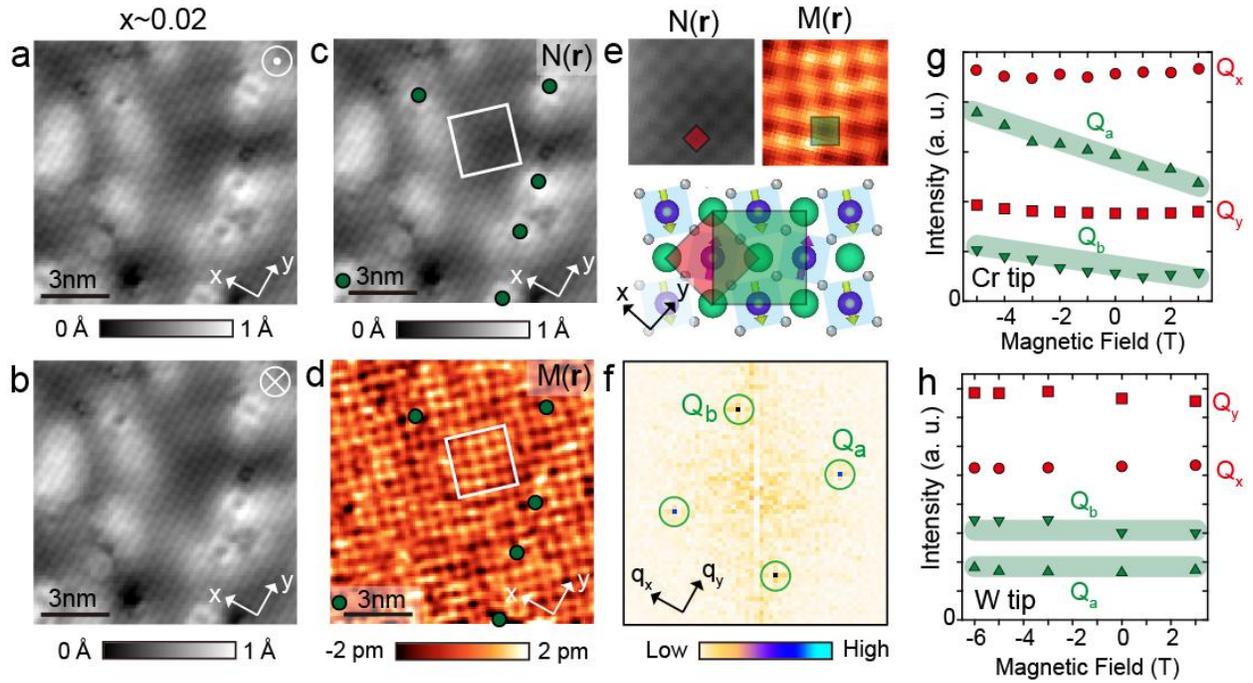

**Figure 2. Spin-resolved magnetic contrast modulations in lightly-doped Sr-214.** STM topographs T(**r**, B) of $x \sim 0.02$ Sr-214 acquired using a spin-polarized Cr tip, in magnetic field (a) B = +3 T and (b) B = -3 T applied perpendicular to the surface (+ (-) sign denotes the field applied parallel (antiparallel) to $\hat{z}$). (c) Arithmetic average of STM topographs in (a) and (b) referred to as N(**r**). (d) Spin-resolved magnetic contrast image M(**r**) obtained by the subtraction of STM topographs in (a) and (b), as defined in the main text. Green circles in (c) and (d) denote the locations of La dopants. (e) Zoom-in on a small region of the sample outlined by a white square in (c,d), showing the atomic-scale structure of N(**r**) and M(**r**), as well as the crystal structure of Sr-214. Pink and green squares in (e) denote the Sr unit cell and the superstructure unit cell, respectively. (f) Fourier transform (FT) of M(**r**) in (d). Intensities of FT peaks extracted from STM topographs T(**r**, B) acquired over an identical area of the sample in varying magnetic field using (g) spin-polarized Cr tip and (h) spin-averaged W tip. For visual purposes, each pixel in (d) is boxcar averaged with the neighboring 8 pixels to smooth over the random noise in the data. We postulate that the difference in the intensity between $\mathbf{Q_x}$ and $\mathbf{Q_y}$ in (g,h) at any given field is a consequence of small tip anisotropy. STM setup conditions: (a,b,g,h) $V_{sample}$= 700 mV, $I_{set}$ = 400 pA.

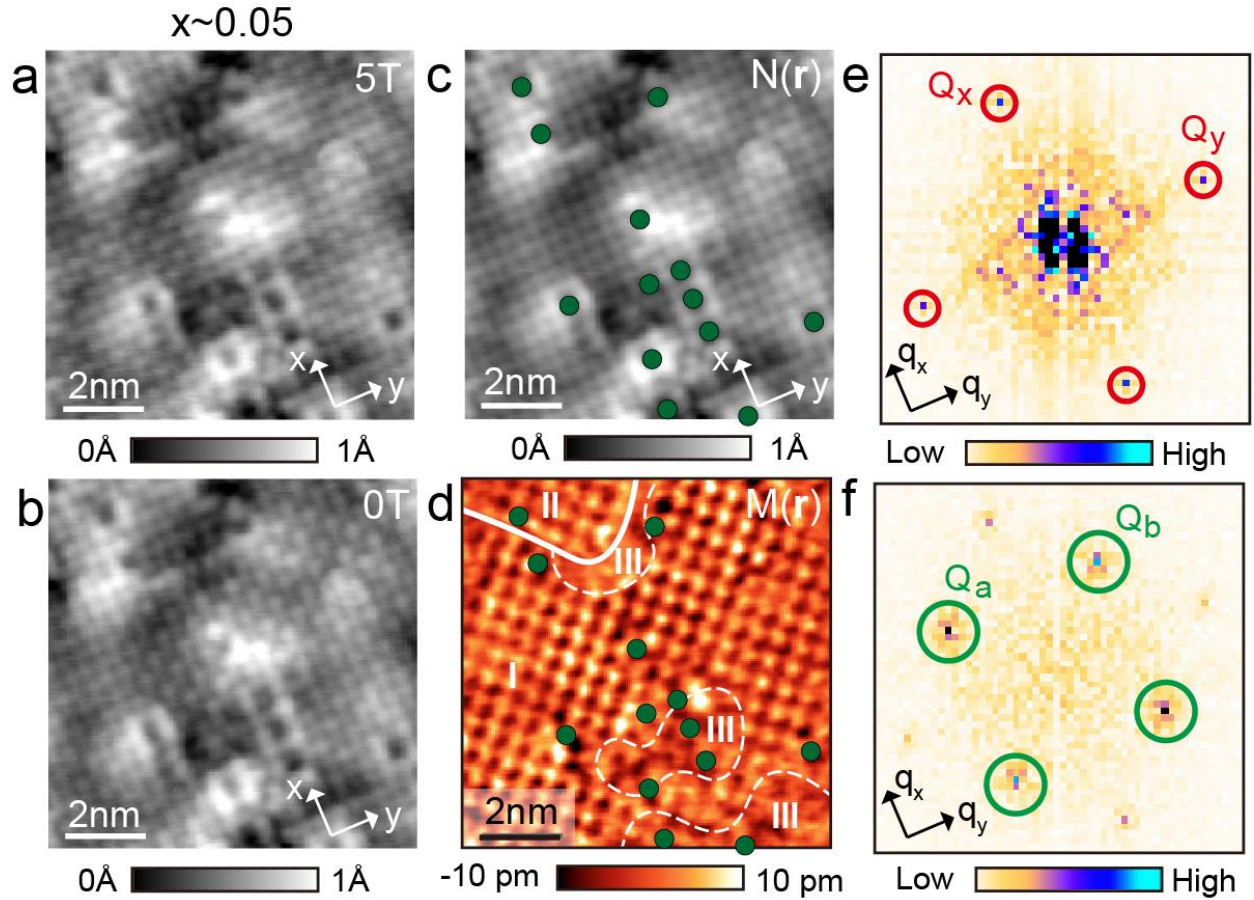

**Figure 3. Fragmentation of the spin-resolved magnetic contrast modulations at higher electron doping.** Scanning tunneling microscopy (STM) topograph T(**r**, B) acquired using a spin-polarized Cr tip and a magnetic field (a) B = 5 T applied perpendicular to the surface and (b) B = 0 T. (c) Arithmetic average of STM topographs in (a) and (b), N(**r**). (d) Spin-resolved magnetic contrast image M(**r**) obtained by the subtraction of images in (a) and (b) as defined in the main text. For visual purposes, image in (d) is separated into 3 different regions: Ordered regions (type I and type II, offset by a phase) and disordered regions (III) (see Fig. S7). (e,f) Fourier transforms (FTs) of images in (c,d), respectively. Relevant FT peaks in (e,f) are denoted by red and green circles. STM setup conditions: (a) $V_{sample}$= 500 mV, $I_{set}$ = 300 pA, B = 5 T; (b) $V_{sample}$= 500 mV, $I_{set}$ = 300 pA, B = 0 T.

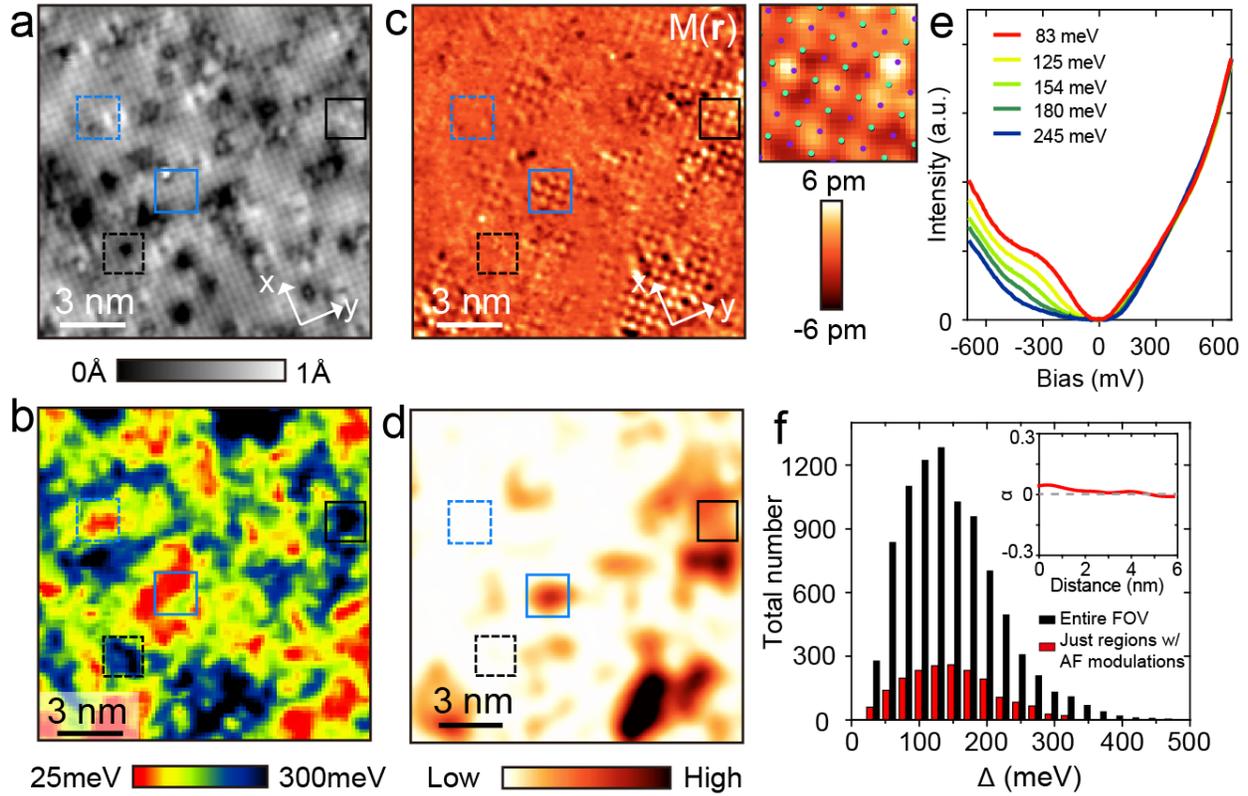

**Figure 4. Relationship between short-range AF modulations and the electronic structure.** (a) Scanning tunneling microscopy (STM) topograph T(**r**, **B**=-0.5 T), (b) map of the spectral gap Δ(**r**) and (c) spin-resolved magnetic contrast image M(**r**), all obtained over an identical region of the sample. Solid black (blue) squares in (a-d) denote regions with spin-resolved modulations and a large (small) gap. Dashed black (blue) squares in (a-d) denote regions *without* spin-resolved modulations and a large (small) gap. (d) Local amplitude map of spin-resolved modulations in (c). The smaller image in (c) is the zoom on the solid black square in (c), with the approximate locations of Sr (green) and Ir (purple) atoms superimposed on top. (e) Average *dI/dV* spectra, binned by the gapmap in (b), showing a systematic evolution with gap magnitude. Legend in (e) shows the average gap magnitudes within each bin, as determined by the algorithm in Ref. (*33*) and described in Supplementary Information 3. (f) Histogram of Δ(**r**) over the entire field-of-view versus the histogram of Δ(**r**) in regions with strong spin-resolved modulations only. Inset in (f) shows angularly-averaged cross-correlation coefficient α between images in (b) and (d). STM setup condition: $V_{sample}$= 700 mV, $I_{set}$ = 1 nA.